# Nonlinear Control of Absorption in Graphene-based 1D Photonic Crystal

M. A. Vincenti[1,*], D. de Ceglia[1], M. Grande[2], A. D'Orazio[2], M. Scalora[3]

[1]*National Research Council - AMRDEC, Charles M. Bowden Research Laboratory, Redstone Arsenal - AL, 35898, USA*
[2]*Dipartimento di Ingegneria Elettrica e dell'Informazione (DIEI), Politecnico di Bari, Via Re David 200, 70126 Bari – Italy*
[3]*Charles M. Bowden Research Laboratory, AMRDEC, US Army RDECOM, Redstone Arsenal - AL, 35898, USA*
[*]*Corresponding author:maria.vincenti@us.army.mil*

**Abstract**: Perfect, narrow-band absorption is achieved in an asymmetric 1D photonic crystal with a monolayer graphene defect. Thanks to the large third order nonlinearity of graphene and field localization in the defect layer we demonstrate the possibility to achieve controllable, saturable absorption for the pump frequency.

Since its experimental discovery [1] graphene has been widely investigated for the richness of its optical and electronic properties, such as charge carriers that mimic massless Dirac fermions, electron-hole symmetry near the charge neutrality point, and weak spin-orbit coupling. Single graphene sheets (0.34nm thick) exhibit universal optical conductivity from visible to near-infrared wavelengths due to inter-band transitions responsible for remarkably high absorption values (~2.3%) [2]. Such peculiar absorption properties may turn graphene into an excellent candidate to replace transparent electrodes or optical display materials [3-5] and a good substitute for commonly used saturable absorbers for laser mode locking [6-10]. Graphene has also been investigated for its ability to support surface plasmons [11, 12], with the aim of promoting light-matter interactions at the nanoscale [13]. In fact, graphene plasmons show significantly more field confinement when compared with their metal counterparts, longer propagation distances, and tuning abilities [11]. The high local fields that accompany surface plasmons on graphene have facilitated the achievement of the near-complete absorption of light in patterned sheets of doped graphene [14]. Enhancement of graphene absorption has been observed when the monolayer is placed on top or within dielectric mirrors [15, 16], or by resorting to surface plasmon excitation [17]. Recent experiments have also demonstrated that single- and multi-layer graphene possess large nonlinear susceptibilities [18-20]. Self-focusing Kerr and saturable absorption coefficients were extracted from z-scan measurements yielding an effective $\chi^{(3)}$ of the order of $10^{-13}$ m$^2$/V$^2$ [21]. On the other hand, third order susceptibility values of the order of $10^{-15}$ and $10^{-16}$ m2/V2 were obtained through four-wave-mixing experiments [22] and third harmonic generation measurements [23], respectively. Even in its lower estimate, the order of magnitude of the nonlinear susceptibility of graphene remains several orders of magnitude larger than most common dielectric materials [24], and comparable to that of metals [25, 26]. Bilayer graphene shows a similar exceptional response for second order nonlinear processes, exhibiting tunable, second order susceptibility as high as $10^{-7}$ m/V [27].

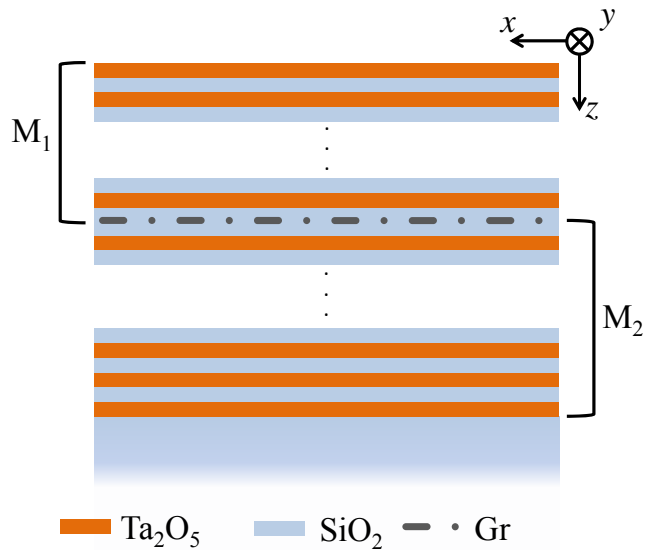

Fig. 1: Sketch of the 1D PhC structure with a monolayer graphene defect layer.

In this paper we propose to exploit the large nonlinear response of a monolayer graphene included in the defect layer of an asymmetric 1D photonic crystal (PhC) to dynamically change the structure from a perfect absorber to a mirror. We show that a defect state forms in the low intensity regime, while the structure's asymmetry leads to narrow-band, tunable, near complete absorption of the pump field. Then, we exploit the third order nonlinearity and saturation effects in the graphene to dynamically suppress absorption by increasing the pump irradiance above a few tens of MW/cm$^2$. Finally, we show that the relatively small field enhancement, introduced by the localized defect state, is nevertheless sufficient to lower the threshold of saturable absorption by nearly two orders of magnitude with respect to bare graphene, and simultaneously increase third harmonic conversion efficiency.

We consider the structure sketched in Fig. 1: the asymmetric quarter-wave 1D PhC is designed to have a transmission defect state at $\lambda_\omega$ = 818 nm, at the center of a transmission band gap. The asymmetric PhC is composed of two mirrors: the top mirror ($M_1$) consists of 6 periods of $Ta_2O_5$ and $SiO_2$, while the bottom mirror ($M_2$) is composed of 15 periods of the same materials, thus forming a $\lambda/2$ $SiO_2$ defect layer. $Ta_2O_5$ and $SiO_2$ layer thicknesses are 96nm and 133nm, respectively. All materials' data are taken from Ref. [28].

A graphene sheet (grey dash-dotted line in Fig.1) is placed at the center of the defect layer. The inclusion of graphene in the defect layer suppresses transmission at the defect state in favor of absorption. In particular, by varying the number of periods of the two mirrors one can tailor the maximum absorption value achieved in the structure. It follows that ~100% absorption can be achieved only thanks to the asymmetry of the PhC with respect to the defect layer.
To understand the benefits of including graphene into the PhC we compare the absorption at normal incidence of bare graphene (black, dotted curve in Fig.2), graphene on top of a single mirror $M_2$ [15] (red, dashed curve in Fig. 2) and graphene placed in the defect layer of a 1D PhC (blue, solid curve – Fig. 2). We note that while placing the graphene on top of a dielectric mirror one can increase absorption over a wide spectral range (red, dashed curve in Fig. 2), a more dramatic enhancement is obtained when the graphene is introduced in the defect state of the asymmetric PhC (blue, solid curve – Fig. 2). The significant improvement of absorption in the latter scenario is attributable to near-order of magnitude electric field enhancement achieved at graphene's position. Absorption as a function of wavelength and angle of incidence for both polarizations is shown in Fig. 2(b) and (c), revealing also the tunable nature of the narrow-band absorption feature.

The graphene sheet is modeled by including an electric current sheet, $J_{x,y} = \sigma_g E_{x,y} d$, on the boundary condition of the tangential component of the magnetic field, where $E_{x,y}$ are the electric field components along the $x$ and $y$ directions, respectively (Fig.1).

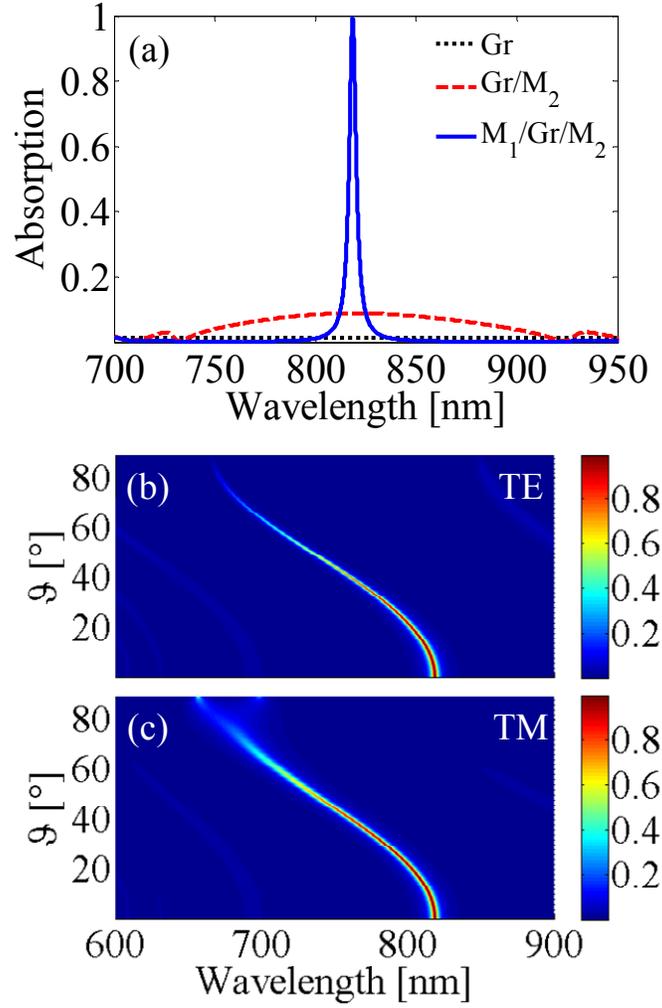

Fig. 2: (a) Absorption as a function of wavelength for bare graphene (black, dotted curve), graphene on mirror $M_2$ (red, dashed curve), and graphene within the defect layer of the 1D PhC in Fig.1 (blue, solid curve); absorption maps as a function of pump wavelength and angle of incidence for (b) TE-polarized and (c) TM-polarized incident field for the structure in Fig.1.

The conductivity value $\sigma_g$ is obtained by using the Kubo formula as in Ref. [29]. The thickness of a single layer of graphene is assumed equal to $d = 0.34$ nm [1]. To account for third order nonlinear effects, saturation effects, and third harmonic generation we express the leading contributions of the nonlinear polarization densities in the j direction at fundamental ($P_{\omega,j}$) and third harmonic ($P_{3\omega,j}$) frequencies as:

$$P_{\omega,j} = 3\varepsilon_0 \sum_{l,m,n=1}^{3} \chi_{jlmn}^{(3)}(\omega,\omega,-\omega,\omega) E_{\omega,l} E_{\omega,m}^* E_{\omega,n}, \qquad (1)$$

$$P_{3\omega,j} = \varepsilon_0 \sum_{l,m,n=1}^{3} \chi_{jlmn}^{(3)}(3\omega,\omega,\omega,\omega) E_{\omega,l} E_{\omega,m} E_{\omega,n}, \qquad (2)$$

where $j,l,m,n$ are the Cartesian coordinates and $\chi_{jlmn}^{(3)}$ are the instantaneous third order susceptibility tensor components. In what follows we consider only the nonlinear polarization at the fundamental frequency (Eq. (1)) assuming that the only non-zero components are $\chi_{xxxx}^{(3)} = \chi_{yyyy}^{(3)} = \chi^{(3)} S$, where $\chi^{(3)} = 10^{-16}(1\text{-}i)$ m$^2$/V$^2$ [23] and the saturation coefficient is [30]:

$$S = \left[1 + \left(|E_x|^2 + |E_y|^2\right) \Big/ |E_{sat}|^2\right]^{-1}. \qquad (3)$$

$E_{sat} = 5.5 \times 10^7$ V/m is assumed to be frequency independent in all simulations. Because of the high third order nonlinearity of graphene one can also expect to observe efficient third harmonic generation processes (see Eq. (2)). An extensive analysis of third harmonic generation is postponed to a future effort. We find that the presence of a localized state in the defect layer improves field localization at the graphene position, and dramatically enhances the nonlinear response of the structure. We calculate absorption as a function of pump irradiance at $\lambda_\omega$ = 818 nm for bare graphene (black, dotted curve in Fig.3), graphene on top of mirror M$_2$ (red, dashed line in Fig.3) and graphene placed within the defect layer of the 1D PhC (blue, solid line – Fig. 3). The figure shows that absorption in the 1D PhC increases 50-fold with respect to a free-standing monolayer. The nonlinear absorption curves are characterized by two different thresholds, which are present regardless of the inclusion of graphene in any resonant structure. The first threshold occurs at lower intensities and indicates where absorption begins to drop thanks to the action of the third order nonlinearity. The second threshold occurs at larger intensities, and indicates where saturation effects dominate preventing absorption losses from vanishing. This effect is taken into account in all three scenarios of Fig. 3 through the saturation factor $S$ (Eq. (3)). When the monolayer is included in the defect layer of the asymmetric PhC both thresholds are lowered by approximately two orders of magnitude with respect to bare graphene. Moreover, while for bare graphene a decrease of absorption corresponds to an increase in transmission, when the graphene is placed either on a single mirror or sandwiched in the defect layer of the 1D PhC the reduction of absorption is paired with increased reflections. More specifically, in the scenario of graphene placed in the defect layer of the 1D PhC dynamic tuning of the structure from perfect absorber to mirror is obtained.

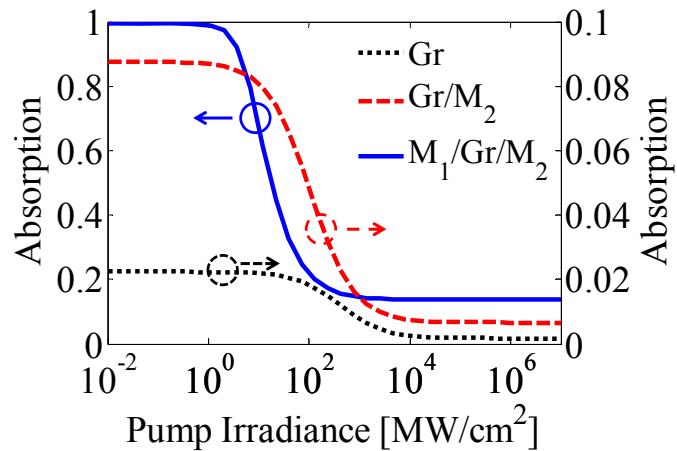

Fig. 3: Absorption as a function of pump irradiance for bare graphene (black, dotted curve), graphene on single mirror (red, dashed curve) and graphene in the defect layer of the 1D PhC in Fig.1 (blue, solid curve).

In conclusion we have shown that the inclusion of a single graphene sheet in the defect layer of an asymmetric 1D PhC structure can lead to perfect, narrow-band absorption. This perfect absorption condition may be realized at any desired frequency by modifying the 1D PhC design and tuned by changing the angle of incidence, regardless of incident polarization. The nonlinear behavior of the structure is rich and complex because of the large third order nonlinearity displayed by graphene, and the presence of saturation effects. The field localization obtained for the graphene included in the asymmetric

PhC suffices to lower the threshold of nonlinear saturable absorption by at least two orders of magnitudes and makes this structure suitable for efficient third harmonic generation.

**Acknowledgement**

This research was performed while the authors M. A. Vincenti and D. de Ceglia held a National Research Council Research Associateship awards at the U. S. Army Aviation and Missile Research Development and Engineering Center. M. Grande thanks the U.S. Army International Technology Center Atlantic for financial support (W911NF-12-1-0292).